\documentclass[letterpaper,twocolumn,10pt]{article}
\usepackage[twoside=true, head=13pt,
     paperwidth=8.5in, paperheight=11in,
     includeheadfoot=false, columnsep=2pc,
     top=1in, bottom=1in, inner=0.75in, outer=0.75in,
     marginparwidth=2pc,heightrounded]{geometry}

\usepackage{graphicx}
\usepackage{booktabs}
\usepackage{url}
\usepackage{algorithm}
\usepackage{algpseudocode}
\usepackage{amsmath}
\usepackage[tt=false, type1=true]{libertine}
\usepackage{mathptmx}
\usepackage{amssymb}
\usepackage[varqu]{zi4}
\usepackage[T1]{fontenc}
\usepackage{enumitem}
\usepackage[font={footnotesize},labelfont={footnotesize,bf},textfont={footnotesize,it}]{caption}
\usepackage{lipsum}

\title{\Large \bf Snow: Self-organizing Broadcast Protocol for Cloud}
\author{Chengkai Tong}
\date{}

\begin{document}
\pagestyle{plain}
\maketitle
\begin{abstract}
Efficient and reliable broadcast protocols are crucial for communication in large-scale distributed systems. Traditional tree-based protocols exhibit rigid topological structures and suffer from performance degradation or even failure under dynamic cluster membership. In contrast, Gossip-based protocols incur high bandwidth overhead and offer only uncertain probabilistic delivery guarantees. Tree-based broadcasting struggles to cope with frequent node dynamics, leading to degraded performance and reliability. However, such issues are intolerable in data center environments. This paper presents Snow, a self-organizing broadcast protocol that is resilient to node change while maintaining high performance. Snow introduces a novel churn-tolerant protocol that guarantees broadcast reliability under dynamic node changes, ensuring that churn does not disrupt stable nodes. To our knowledge, this work presents the first formal model for churn resilience in this setting. Compared to baselines, Snow achieves higher reliability and lower traffic overhead while reducing latency by 30\%. Building upon Snow, we designed the Coloring algorithm, which leverages idle bandwidth at leaf nodes to double the message convergence speed.
\end{abstract}
\section{Introduction}
Modern large-scale cloud applications—including real-time communication systems like video conferencing and instant messaging(IM) platforms \cite{saint2011extensible}—require efficient data dissemination mechanisms. Broadcast algorithms serve as a powerful mechanism for accelerating tasks such as pushing configuration data \cite{hunt2010zookeeper}, distributing MapReduce\cite{dean2008mapreduce} or Spark \cite{zaharia2012resilient} binaries, replicating blocks in GFS \cite{ghemawat2003google} and HDFS \cite {borthakur2008hdfs}, and delivering messages in message queues \cite{kreps2011kafka}. Utilizing broadcast-based dissemination alleviates traffic congestion at the source by distributing relay tasks to downstream child nodes, thereby distributing the forwarding load more evenly across the network. This approach accelerates data distribution while avoiding a bottleneck at a single node. Another example is the publish-subscribe system \cite{eugster2003many,armbrust2009above}: On social media platforms, each user subscribes to others whose content they find interesting. When users with massive followings post new updates, this information needs to be delivered to millions of followers. From the user's perspective, efficient and reliable transmission of subscribed content is deemed critical, which presents a fundamental challenge in designing systems that maintain delivery guarantees across asymmetric follower relationships. 

Dynamic resource allocation is standard practice in cloud environments \cite{armbrust2009above}, as cloud services operate on a usage-based billing model \cite{elmroth2009accounting}. To optimize costs, enterprises must continually adjust server capacity in response to fluctuating workloads \cite{xiao2012dynamic,xiao2012automatic,zhang2021sinan,qiu2023aware}. In this context, an efficient broadcast algorithm becomes particularly significant. An ideal broadcast algorithm should handle frequent node changes without message loss while minimizing bandwidth consumption—with fixed fan-in and fan-out bandwidth constraints, lower broadcast traffic overhead enables each node to support more users, thereby reducing operational costs for enterprises. 

However, there are notable differences between the assumption of existing distributed broadcast algorithms and the backend infrastructure. In data center networks, the Fat-trees\cite{al2008scalable,leiserson1985fat} topology interconnects distinct nodes via switch trunks, enabling direct communication between any pair of nodes without requiring intermediate forwarding through other nodes. Based on real-world observations\cite{itic2022}, cloud cluster states typically fall into three scenarios, ordered from least to most common. 1) Unexpected Crashes: Node failures due to hardware or system issues are rare. 2) Nodes go up/down as expected. 3) All nodes staying perfectly unchanged is common. The above observations guide the design of our protocol. In addition to that, most cloud service providers do not offer built-in support for IP multicast \cite{deering1990multicast,li2013scaling}. An alternative to IP multicast is application-level multicast, as it operates independently of specialized network hardware and can be widely implemented, offering greater flexibility and broader applicability. 
% However, this method increased transmission overhead owing to the absence of IP multicast capability, as they must resort to unicast transmission to each recipient individually to achieve broadcast. 
Therefore, our method \textbf{Snow} adopts a purely application-level approach to enable broadcasting among a vast number of entities in a network.

In this paper, we propose a novel protocol Snow: a self-organizing distributed broadcast protocol for the cloud environment. In a decentralized service architecture, disseminating node state information through broadcast mechanisms enables the service cluster to achieve self-organization \cite{lakshman2010cassandra,van1998Gossip}. In contrast to prior approaches \cite{li2013reliable,li2011esm,iyer2014avalanche}, Snow operates exclusively through mutual node communication, eliminating the need for specialized hardware support.

Leveraging the small amount of payload information attached to each message, Snow dynamically selects nodes from the membership list as child nodes to broadcast the message. In addition, frequent changes in nodes do not affect the performance and stability of the cluster. The final broadcast topology of Snow is similar to a balanced multi-way tree construct with the source node as the root. Multiway balanced trees exhibit uniform leaf depths with at most one-level variation, ensuring all root-to-leaf paths differ by at most one hop and broadcast messages reach all nodes in $\mathrm{O}(\log n)$ hops. Unlike traditional tree broadcasting protocols \cite{lim2000multicast,jannotti2000overcast}, which require building a tree for each source to determine their next hop in the cluster, Snow only needs to maintain a list of IP addresses, and all nodes are able to automatically route according to the appropriate path. Furthermore, a node's departure under non-faulty conditions does not disrupt ongoing communication between remaining nodes. Snow automatically updates the network topology without intervention in such instances. When a node fails due to unexpected reasons, such as hardware issues, Snow detects it within seconds and evicts the faulty node from the cluster via its built-in broadcast mechanism, minimizing broadcast message loss.

Crucially, different from prior work, Snow ensures that membership changes never disrupt message delivery to nodes in the cluster. This guarantees that nodes in the stable portion of the cluster remain entirely unaffected, even under extreme churn. We argue that this is profound: Consider a scenario where master-slave cache synchronization \cite{redis_arch} is performed via broadcast. This feature is particularly appealing, as it enables the algorithm to remove a slave node without disrupting the normal message synchronization among the remaining slave nodes. To our knowledge, Snow is the first broadcast protocol to propose such a churn-resilient model. 

In summary, we make the following contributions:
\begin{enumerate}
\item We propose a message broadcasting protocol for the cloud, capable of reaching any node in the cluster after the message is distributed $\mathrm{O}(\log n)$ times.
\item We introduce a service discovery mechanism based on Snow, which provides reliable membership maintenance for large-scale clusters.
\item  We design a novel algorithm that allows frequent changes in cluster membership without compromising the efficiency and reliability of the broadcast algorithm.
\end{enumerate}

\begin{table*}[htbp]
    \centering
    \label{tab:protocol_comparison_transposed}
    \begin{tabular}{lccccc}
        \toprule
        \textbf{Scene} & \textbf{Node Failure} & \textbf{No Failures} & \textbf{Stragglers} & \textbf{Node Changes} & \textbf{Traffic} \\
        \midrule
        Gossip & Probabilistic delivery & Probabilistic guarantee & Robust & Unaffected & High \\
        Tree-based & Delivery failure possible & Guaranteed delivery & Vulnerable & Affected & Low \\
        \textbf{Snow} & Ensures critical message delivery & Guaranteed delivery & Mitigated & Unaffected & Low \\
        \bottomrule
    \end{tabular}
    \caption{Comparison of Protocol Characteristics}
    \label{tab:compare}
\end{table*}
\section{Problem Statement}

This paper focuses on broadcasting within a single cluster inside the same data center. We summarize several key challenges in data center broadcasting.

\textbf{Bandwidth Limitation:}
We define the number of channels through which a node receives messages as fan-in, and the number of channels through which a node sends messages as fan-out. Furthermore, when a message is accepted by all nodes, we say that the message has converged. Each node has multiple fan-out connections to broadcast messages rapidly to all nodes. Increasing the number of nodes' fan-out could result in faster convergence. This also led to fan-out becoming a resource bottleneck, and some nodes that receive messages last are unable to efficiently utilize their fan-out bandwidth. During large file broadcasting, this issue becomes particularly critical—the substantial traffic overhead not only impacts network efficiency but also induces significant CPU and memory pressure.

\textbf{Straggler Problem:}
In a cluster, different nodes handle varying workloads, and a high load could significantly degrade the quality of service (QoS), causing some nodes to become stragglers \cite{ouyang2016straggler,ananthanarayanan2014grass}. In a typical cluster environment, only a small subset of nodes become stragglers, as the workload and hardware conditions of the majority of nodes have an insignificant impact on program execution and broadcast performance . The set of stragglers might change over time due to variations in the load of each node. A straightforward approach is to establish multiple disjoint message paths for each node; however, this measure may inadvertently exacerbate bandwidth consumption at the nodes.

\textbf{Membership Change:}
In a clustered environment, node joining and departure are common. A typical scenario involves the dynamic adjustment of the number of nodes in a cloud infrastructure in response to fluctuating workloads, but the silent departure of nodes due to hardware failure or system issues is relatively rare \cite{itic2022}. Specifically, most nodes must complete designated tasks before going offline in order to perform a graceful shutdown, thereby minimizing the impact on the cluster. For many dissemination topologies, the joining and leaving of nodes seriously impact the stability and efficiency of message broadcasting within the cluster. However, nodes in a cluster within a data center are strictly controlled when it comes to coming online and going offline.

\section{Related Work}
\label{sec:Existing Protocols}

\textbf{Flooding \cite{lim2001flooding}:} The node broadcasts data to all the neighboring nodes. This could result in excessive fan-out of a single node within a certain period of time. The flooding mechanism is straightforward to implement, making it particularly well-suited for deployment in small-scale clusters and scenarios characterized by low data volumes. In large-scale systems with numerous nodes or high data volumes, this can lead to pronounced performance degradation and considerable latency overhead. 

\textbf{Tree-based Broadcasting:} Tree broadcasting has various applications in different network structures\cite{ganguly2005fast,juttner2005tree,chlamtac1987tree,ganesh2001scamp,junhai2008research,chu2002case}, where messages are forwarded from parent nodes to child nodes along the tree topology. Tree-based broadcasting is fragile, and the failure of the parent node leads to the message being unreachable. Furthermore, every time a node joins or leaves, the tree structure needs to be adjusted. Even though we built a tree that minimizes the average cost and enables the fastest distribution of messages to all nodes within the cluster. However, from the perspective of all other nodes in the tree except the root, this tree is not optimal for message broadcasting. A common approach is to maintain a tree for each root node (a node that initiates a broadcast). However, this way not only increases memory consumption but also adds complexity to tree adjustments. In the absence of faults, the tree-based algorithm ensures the delivery of messages, but if a straggler appears on a node in the tree, it drastically degrades the algorithm's performance. Unlike tree-based broadcast protocols, Snow does not require nodes to maintain parent-child relationships. Instead, each node dynamically selects forward targets upon receiving a message, guided solely by the information carried within the message itself.

\textbf{Gossip:} Many broadcasting algorithms are based on Gossip\cite{deshpande2006crew,birman1999bimodal,gupta2002efficient,eugster2003lightweight}. A classic approach is this: When a node receives a message for the first time, it selects $k$ distinct nodes as fan-out at random from the set of all reachable nodes and forwards the message to them. The forwarding process continues iteratively until the message reaches all active nodes in the network (perhaps never). As the clusters expand, in order to achieve atomic broadcasting \cite{kermarrec_probabilistic_2003} (all nodes receive the message) with a higher possibility, so that every node is guaranteed to receive the notification, it is necessary to maintain $k \geq \log(n)$ \cite{pereira2001probabilistic}, which leads to an increase in node load. When $k=n-1$, Gossip degenerates into flooding. Nevertheless, even if $k$ is set to $n-2$, it is still possible that atomic broadcasting cannot be achieved in the absence of cluster failures—specifically, all nodes coincidentally miss the same node. Some algorithms, such as HyParView \cite{leitao2007hyparview}, reduce the number of memberships each node needs to maintain to $\mathrm{O}\log(n)$ nodes. The cost is additional communication and complexity in maintaining partial views, which is deemed unnecessary in a daily scenario. 

The differences between Snow and the aforementioned protocols are summarized in Table \ref{tab:compare}.

\textbf{Other Work:}
Snow different from some existing broadcast protocols \cite{castro2002scribe,ratnasamy2001application,alibaba_dragonfly_2018,uber_kraken,zhuang2001bayeux}, they adopt peer-to-peer infrastructure\cite{rowstron2001pastry,ratnasamy2001scalable,cohen2003incentives,zhao2001tapestry}. These infrastructures are specifically designed for adversarial environments, where the likelihood of nodes leaving abnormally and encountering failures is substantially higher than in a stable data center environment. In real data centers, the latency difference between nodes is negligible. Furthermore, unlike the original intention of peer-to-peer infrastructure mentioned above, in data centers, nodes directly send messages to each other without the need for forwarding from other nodes.

Several studies have explored broadcasting across data centers \cite{cohen2003incentives,luo2019deadline,zhang2018bds,tseng2021codedbulk}, and their approaches are complementary to ours. These works navigate different trade-offs among cost, speed, and reliability when transmitting data across geographically distributed data centers, primarily due to the substantial cost disparity of inter-data center communication. Consequently, transmission mechanisms in these environments require different designs compared to intra-data-center optimized algorithms.

Plumtree\cite{leitao2007epidemic} combines the advantages of tree-based and Gossip-based broadcasting. It first constructs a spanning tree using the nodes initially reached via Gossip. Under stable cluster conditions, messages are distributed along this tree. When cluster membership changes, Gossip is reactivated to reconnect the affected parts of the tree. However, this topology has biases regarding the state of the cluster, and as more nodes join, the tree becomes increasingly inaccurate. Both the randomness in initial broadcast and dynamic node membership can trigger tree reconstruction, increasing protocol overhead. This behavior is unacceptable for real-time distributed systems, particularly in scenarios requiring low-latency broadcast operations such as cache coherence.

SWIM\cite{das2002swim} is a protocol for maintaining membership lists in the cluster, which is able to support larger clusters with lower costs. In SWIM, both node joining and voluntarily leaving require the use of Gossip broadcasting messages. When a node leaves due to a fault, it will be detected by the heartbeat of other nodes. The work of Lifeguard\cite{dadgar2018lifeguard} is based on SWIM, which adopts a more cautious and reasonable mechanism to reduce false positive rates of detection. This mechanism has been operating efficiently in a cluster with over 6,000 nodes. Our work on node health checks in Snow is based on SWIM and Lifeguard, but we used our own broadcast mechanism instead of the original Gossip to update the status of nodes.

\section{Design}
\subsection{Overview}
Snow is a decentralized broadcast protocol where all nodes within a cluster are peers, and each node is capable of initiating broadcasts or forwarding arbitrary messages. In decentralized service architectures, by broadcasting the state of a node to other nodes \cite{lakshman2010cassandra}, the service cluster achieves self-organization, maintaining awareness of the liveliness of member nodes without requiring consensus protocols or additional middleware. In Snow, the member organization and broadcast algorithms are interdependent, specifically manifested in that each broadcast operation requires computation based on the internal membership of nodes, while updating the member view relies on the broadcast algorithm. Since the final topology forms a balanced multi-way tree, for simplicity, we continue to describe the broadcast process using tree terminology in the following section.

\begin{algorithm}
\caption{Algorithm for determining nodes}
\label{alg:FindNode}
\begin{algorithmic}[1]
\State {FindNode$(left, right, current, k)$:}
\State ${init}$ ${childNodes} \gets {empty list[boundary, node, boundary]}$
\State $k' \gets \lfloor k / 2 \rfloor$
\If{${right} - {left} \leq k$}
    \For{$i = {left}$ to ${right}$}
        \If{$i = {current}$}
            \State \textbf{continue}
        \EndIf
        \State ${childNodes.add(i, i, i)}$
    \EndFor
    \State \Return ${childNodes}$
\EndIf
\State ${rightRegionSize} \gets \lfloor ({right} - {current}) / k' \rfloor$
\State ${lBoundary} \gets {current} + 1$
\For{$i = 0$ to $k' - 1$}
    \State ${rBoundary} \gets {lBoundary} + {rightRegionSize} - 1$
    \State ${mid} \gets \lfloor ({lBoundary} + ({rBoundary} + 1)) / 2 \rfloor$
    \State ${childNodes.add}$ $({lBoundary}, {mid}, {rBoundary})$ {}
    \State ${lBoundary} \gets {rBoundary} + 1$
\EndFor
\State ... Use the same method on the left
\State \Return ${childNodes}$
\end{algorithmic}
\end{algorithm}

\vspace{-10pt}
\subsection{Standard Snow Broadcasting Mechanism}
In the broadcasting process, we denote $k$ as the fan-out degree of each node, i.e., the number of child nodes per node. Among them, any multiple of 2 is allowed for $k$; It is a parameter defined manually and remains immutable once the cluster has been initialized. We refer to nodes that do not forward received messages as leaf nodes, and other nodes as internal nodes.

\subsubsection{State Initialization}
\label{sec:State Initialization}
In the initial setup, Snow discovers the IP addresses of other nodes through configuration. These IPs are sorted and stored in an array. Any changes to the node membership will trigger an update to this array while maintaining its sorted order. 
% Due to the principle of locality in arrays, this operation incurs a negligible overhead. 
If uniformity is necessary, the IP address and port can be hashed using a function such as BLAKE2 \cite{aumasson2013blake2} or SipHash \cite{aumasson2012siphash}, and the resulting hash values serve as identifiers. For simplicity, the following context only describes implementation directly using the IP address and port. Unlike finger-table of Chord\cite{stoica2001chord}, Snow adopts a full membership view. However, the amount of space consumed by this IP address is small. With an IPv6 address plus 2 bytes for the port, each node occupies only 18 bytes. Thus, 100,000 nodes require only under 2 megabytes of memory. The membership list remains consistent across all nodes, except during node join and leave events, which will be discussed in detail in Section \ref{sec:Membership Maintenance}.

The node array can be abstracted as a logical ring topology. This symmetry allows Snow to initiate broadcast operations from an arbitrary node on the ring. Here, the root node is denoted as $N_0$. We denote the number of nodes in the cluster as $n$. The nodes to the left and right of the root node are $N_{n-1}$ and $N_1$, respectively, where $N_{n}=N_0$.

\subsubsection{Root Node Broadcasting}

\begin{figure}
    \centering
    \includegraphics[width=1\linewidth]{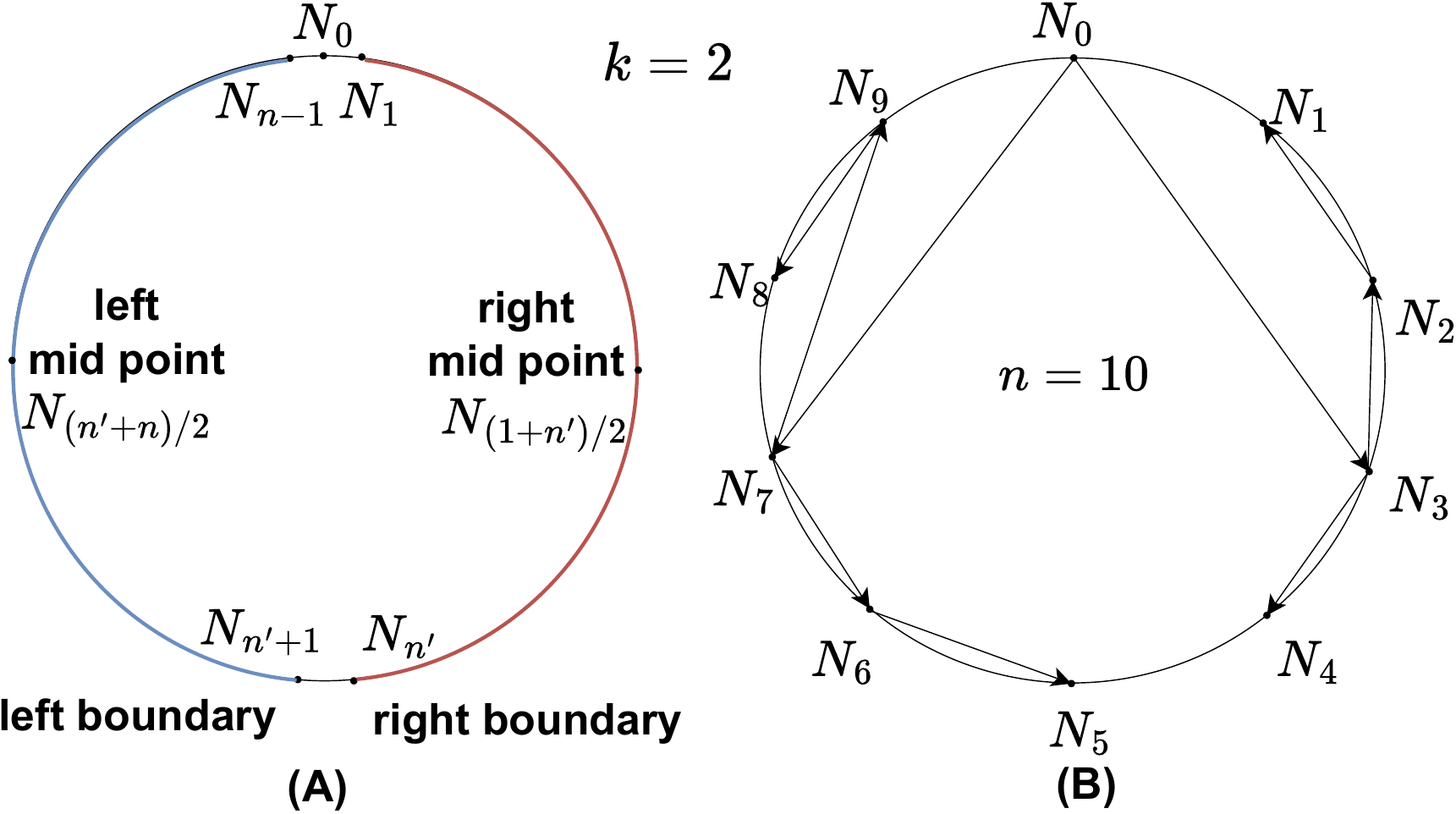}
    \caption{For A, the left region is a blue line, and the right region is a red line. Figure B shows the case where $n=10$.}
    \label{fig:ring}
\end{figure}

At the beginning, a node needs to start broadcasting messages to all other nodes in the cluster; We refer to this node as the root node. Snow first finds the left and right boundaries of the node.
\begin{equation}
[N_1,N_{n-1}]
\end{equation}
 As the left and right boundaries are in charge of the same size of regions, the size of each allocated region is $n'=\frac{n-1}{2}$. Similarly to binary search, when the number of nodes is not divisible, choosing either the left or the right node is feasible. In our default implementation, we choose the right node. In other words, if the number of nodes cannot be evenly divided,  the left region gets one more node than the right. The boundaries of these two regions are: 
\begin{equation}
\begin{split}
\text{rightBoundary} & : N_{n'} \\
\text{leftBoundary} & : N_{n^{\prime}+1}
\end{split}
\end{equation}

After determining the boundaries, Algorithm \ref{alg:FindNode} is ready for execution. Divide $k$ into two parts evenly and uniformly assign them to the two regions, such that each region is allocated $k'=k/2$. Since $k$ is a multiple of 2, $k$ must be divisible by two, as shown in Figure \ref{fig:ring}A. Owing to the inherent properties of the ring structure, the current node is also regarded as the intermediate node between the two regions, as follows (line 13 of the Algorithm \ref{alg:FindNode}):
\begin{equation}
\begin{aligned}
\text{leftRegion} & : [N_{n'+1}, N_{n-1}] \\
\text{rightRegion} & : [N_1, N_{n'}]
\end{aligned}
\end{equation}

At this time, the node is split into three parts: the left region, the right region, and the root node. These two regions will eventually be divided into $k'$ equal regions respectively: 
\begin{equation}
\begin{aligned}
\text{leftRegion} & :\scriptstyle \left\{ [N_{n'+1}, N_{n'+1+\frac{n'}{k'}}], \dots, [N_{n'+1+\frac{(k'-1)n'}{k'}}, N_{n-1}] \right\} \\
\text{rightRegion} & :\scriptstyle \left\{ [N_1, N_{\frac{n'}{k'}}], [N_{\frac{n'}{k'}+1}, N_{\frac{2n'}{k'}}], \dots, [N_{\frac{(k'-1)n'}{k'}+1}, N_{n'}] \right\}
\end{aligned}
\end{equation}

Due to the structure of the ring, the root node always considers itself as the midpoint between the left and right regions. Therefore, Figure \ref{fig:ring}B and Figure \ref{fig:snow-standard} are structurally equivalent. The numbers below the node depict the left and right boundaries of the node, and in practical implementation, we set the IP address and port of the nodes as boundaries.
As the method is identical for both sides, we focus on the right region for clarity. We use the endpoints in the region as the boundaries of this region.

\begin{equation}
\{\{N_1,N_{\frac{n'}{k'}}\},\{N_{\frac{2n'}{k'}+1},N_{\frac{3n'}{k'}}\},...,\{ N_{\frac{(k'-1)n'}{k'}+1},N_{n'-1}\}\}
\end{equation}

Then, Snow finds the midpoint node for each boundary group. This process occurs in lines 15-20 of Algorithm \ref{alg:FindNode}.

\begin{equation}
\begin{split}
\{N_{(1+\frac{n'}{k'})/2},N_{(\frac{n'}{k'}+\frac{2n'}{k'}+1)/2},...,N_{(\frac{(k'-1)n'}{k'}+N_{n'})/2}\}\\
=\{N_{(\frac{n'}{k'}+1)/2},N_{(\frac{3n'}{k'}+1)/2},...,N_{\frac{(2k'-1)n'}{k'}/2}\}
\end{split}
\end{equation}
Use these $k$ points as candidate nodes for the broadcast, which are the child nodes of the current node.
The root node sends the message, along with the left and right boundaries, to these $k$ nodes. The purpose of conveying two boundaries in a message is to inform the child nodes of the regions they need to be responsible for. If the number of nodes within these two regions is less than or equal to $k$, both the left and right boundaries are configured to the candidate node's own IP address, as shown in Figure \ref{fig:snow-standard}.

\subsubsection{Child Nodes Broadcasting}
\label{sec:Second step}
When a new node receives a message, it parses the left and right boundaries in the message and finds them in its own membership list. Due to the fact that the membership list might not be completely synchronized, if the boundary nodes are not found in the membership list, the IP and ports of the nodes will be inserted into the list. Now, every node that receives the message is assigned new responsible regions. As each node performs the same operation, let's take node $N_{(\frac{n'}{k'}+1)/2}$ as an example. Based on two boundaries, we are able to calculate the regions just like the root node:

\begin{equation}
\begin{aligned}
\text{leftRegion} & : \left[N_1, N_{\left(\frac{n'}{k'}+1\right)/2 - 1}\right] \\
\text{rightRegion} & : \left[N_{\left(\frac{n'}{k'}+1\right)/2 + 1}, N_{\frac{n'}{k'}+1}\right]
\end{aligned}
\end{equation}

\begin{figure}
    \centering
    \includegraphics[width=1\linewidth]{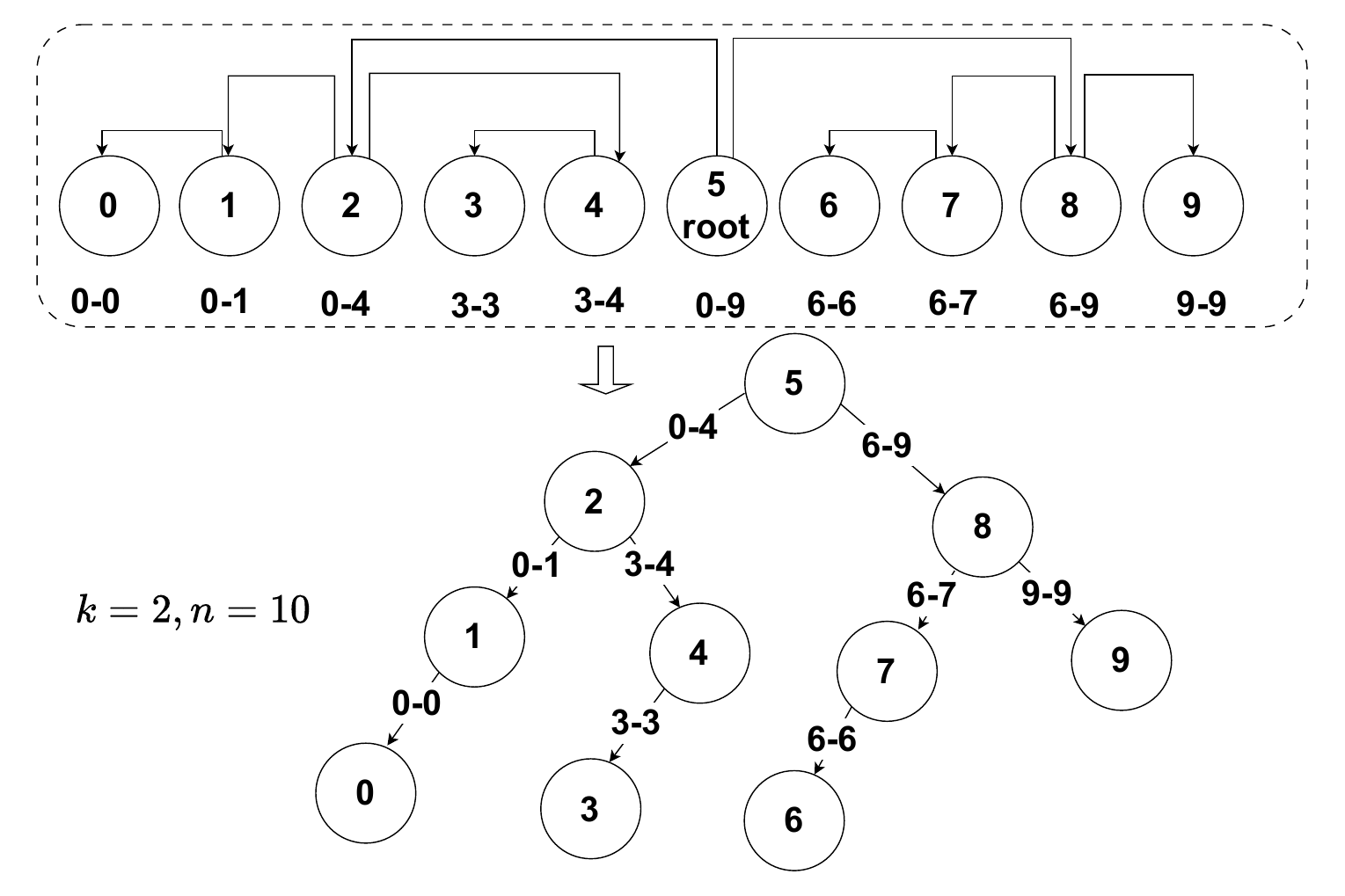}
    \caption{Illustration of the message dissemination process when node 5 initiates a broadcast. In the figure, all nodes have the same and correct membership view.}

    \label{fig:snow-standard}
\end{figure}

The only difference from the root node is that the root node must compute the left and right boundaries by itself. The subsequent nodes do not need to perform this operation, as their boundaries have already been determined by their parent nodes. All internal nodes can be regarded as the root nodes of their respective subtrees, as they serve as the starting points of these subtrees. In other words, the left and right halves of this region form a logical ring again. Consequently, Snow is able to iterate the First step again, as shown in Figure \ref{fig:snow-standard}. When $k=2$, this operation is similar to performing a binary search in both regions for which the node is responsible \cite{el2003efficient}. Notably, due to the introduction of explicit boundary limitations, the size of the regions is reduced in every iteration. When the number of nodes in the region becomes less than or equal to $k$, these nodes are designated as the leaf nodes of the tree (4-12 lines of the Algorithm \ref{alg:FindNode}).

\subsection{Algorithm Review}
\label{sec:Algorithm Review}

The aforementioned algorithm appears to merely construct a balanced tree, but since each node's decision is dynamically made based on the region it obtains, in cases of drastic node changes, each node may maintain a unique membership list $S$. Based on Snow's design above, we can prove that, under the assumption of no message delivery failures or node crashes, for any node $N_x$, if $\forall S, \, N_x \in S$, then $N_x$ is guaranteed to receive the message. Proven in Appendix \ref{appendix:Convergence Proof}. In Appendix \ref{appendix:diff}, we describe an example where each node has a different Membership list. In this case, node $N_x$ can still receive messages. Snow's design ensures that the QoS of stable nodes remains unaffected during cluster changes. We highlight that the delivery guarantee is critical. Consider video conferencing, where the join/leave events of some participants must not cause packet loss or increased latency for others. 

\textbf{Why is this algorithm not tree-based}: Although tree terminology is used to describe the algorithm's execution process, our algorithm is not tree-based. In reality, the tree structure is drawn by connecting the paths traversed by message broadcasts to illustrate the propagation process of the message. In any broadcast protocol, we connect all nodes based on message propagation, keeping only the first incoming link through which each node initially received the message; the resulting structure must be a tree, even in Gossip-based dissemination. In Snow, each node operates myopically, only aware of its designated forwarding target. The ordered membership list in Snow is maintained to ensure that nodes within the region can be found, in order to establish a robust model, rather than to generate a search tree for broadcasting. Consequently, the propagation paths generated by this mechanism structurally resemble a multi-way balanced tree, yet the mechanism itself is not tree-based.

\subsection{Reliable Message}
\label{sec:Reliable Message}

While Snow utilizes a heartbeat mechanism for periodic failure detection and node removal, delayed removal of failed nodes inevitably leads to broadcast unreachability. We will discuss this issue in detail in Section \ref{sec:Membership Maintenance}.

For certain critical messages, Snow ensures reliable broadcast to all non-faulty nodes even in the presence of node failures. Such messages are termed Reliable Messages. In the case where membership lists become inconsistent across nodes, as mentioned above, those nodes that are recognized by all members will invariably receive broadcast messages. This class of messages usually requires determining the convergence time, such as the distribution of container images \cite{verma2015large}, and the user wants to know the completion time of the operation. 

\begin{figure}
    \centering
    \includegraphics[width=0.5\linewidth]{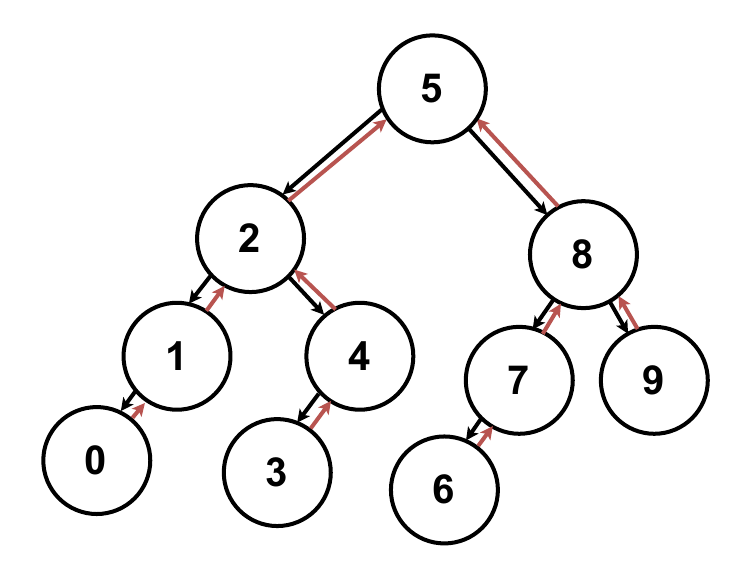}
    \caption{Reliable Message, The black line is a normal message broadcast, and the red line is the acknowledgment.}
    \label{fig:reliable}
    \vspace{-10pt}
\end{figure}

The design of Reliable Message is as follows: when a leaf node receives a message, it sends an acknowledgment directly to its parent node. When an intermediate node has received acknowledgments from all its child nodes, it also sends an acknowledgment to its own parent. As illustrated in Figure \ref{fig:reliable}, black lines represent message propagation, while red lines denote acknowledgments, which propagate from the leaf nodes to the root node. The acknowledgment only needs to contain the message ID, thus consuming minimal bandwidth. In the absence of failures, the convergence time of Reliable Message aligns with that of the standard Snow algorithm, as acknowledgment messages do not affect convergence latency, though they introduce a modest increase in bandwidth overhead.

It is evident that any node failure must prevent the root node from receiving all acknowledgments. Unlike the ACK mechanism in TCP, which directly responds with an ACK after receiving a message, the acknowledgment in Reliable Message is sent to the parent node only after the current node has ensured that all its child nodes have successfully received the message. Once the root node receives a sufficient number ($k$) of acknowledgments, it indicates that no failures occurred during the broadcast and the message has been fully disseminated. However, in cases where the cluster state is inconsistent, some unstable nodes might still miss the message, a scenario that we will discuss in Section \ref{sec:Membership Maintenance}. If the node misses some acknowledgments after a certain time window, it triggers a timeout mechanism and rebroadcasts the message. This time window is usually sufficient to remove the faulty nodes, so the cluster returns to a normal state during the rebroadcast.

\subsection{Membership Maintenance}
\label{sec:Membership Maintenance}
% Snow is self-organizing and also supports maintaining a membership list through external services, such as Zookeeper\cite{hunt2010zookeeper}. A detailed investigation of this topic is beyond the scope of this paper. Our discussion focuses exclusively on the self-organizing aspects of Snow's architecture. 
In the self-organizing component of Snow, each membership change is broadcast as a Reliable Message to notify other nodes to update their membership list.

\subsubsection{Node Join}
\label{sec:Node join}
When a new node joins, it first establishes a connection with an arbitrary existing node in the cluster to synchronize the membership state. Once the synchronization is complete, the new node can start forwarding data. At this point, the new node broadcasts a join request, asking other nodes to add it to their membership list. In Snow, each time a normal node joins or leaves, a broadcast is triggered to inform other nodes to update their membership view. To ensure the stability of the cluster during the addition of new nodes, Snow made practical trade-offs that align with real-world requirements. Specifically, this means that existing nodes in the cluster do not miss any messages, whereas newly added nodes are at risk of missing part or all of the messages broadcast in the system when they join the cluster.

\textbf{Message Loss Prevention:} Suppose each node maintains a membership view set \( S \), and initially, all nodes maintain the same node set \( S_p \) before the new node is added. When a new node \( N_j \) requests to join the set, it first establishes a connection with an arbitrary node and synchronizes all the data of that node. Afterward, node \( N_j \) adds itself to the membership view, meaning that at this point, node \( N_j \) maintains \( S_p \cup \{N_j\} \). We define the latest membership view \( S_l = S_p \cup \{N_j\} \), where \( S_p \subset S_l \). After completing the synchronization, node \( N_j \) broadcasts its join request to all the nodes. During the process of broadcasting, some nodes' current views are \( S_p \), while others are \( S_l \). However, for all the sets \( S \) maintained by the nodes, it holds that \( S \supseteq S_p \). This means that all nodes have the same knowledge about the nodes in \( S_p \). As discussed in Section \ref{sec:Algorithm Review}, any node that is known to all participating nodes is guaranteed to receive the message; The nodes in \( S_p \) miss no messages. In other words, except for the newly joined node \( N_j \), no node lose any messages.

When multiple nodes leave and join the cluster simultaneously, these nodes may miss new member changes. Similarly to Lifeguard \cite{dadgar2018lifeguard}, Snow periodically synchronizes the membership list of a randomly selected node. The mechanism, termed \textit{anti-entropy}\cite{demers1987epidemic}, is executed every 15 seconds by default. \textit{Anti-entropy} guarantees that, even in cases where the join message fails to be successfully broadcast, all nodes will ultimately converge to the correct state of the entire cluster. From the moment all nodes in the cluster become aware of the new node, it will not miss any messages with the same reliability guarantees as existing nodes. As quantified in Section \ref{sec:State Initialization}, the global cluster state incurs only a negligible memory footprint, posing an insignificant impact on datacenter network resources. 

\subsubsection{Normal Leave}
\label{sec:Normal leave}
Similarly, before a node leaves, all nodes have an accurate view of the current cluster state. Snow enforces nodes to perform graceful departure procedures when leaving the cluster. Before leaving the cluster, nodes send a leave request when they intend to leave the cluster. Similar to when nodes join, before message convergence, all nodes have consistent knowledge of the previous nodes, and there might be differences in the state of the departing nodes. Therefore, except for the node that is currently preparing to leave, other nodes do not lose messages. Once the leave message has fully propagated, the departing node disconnects from other nodes after a few minutes of waiting. During this brief waiting period, nodes must still forward messages as usual. By the time a Reliable Message acknowledgment is received, the stable portion of the cluster has already removed departing nodes. The multi-minute window allows the \textit{anti-entropy} mechanism to propagate, ensuring that nodes that missed updates (such as newly added nodes) can still synchronize the correct membership list.

Ultimately, all remaining nodes have an up-to-date view of the cluster state, making sure that no messages are lost. In summary, whether a node joins or leaves, it does not affect the stable nodes in receiving messages, nor requires sending redundant messages, which would create additional load. Most importantly, the message convergence time does not increase. When multiple nodes join and leave simultaneously, the stable portion of the cluster can always receive messages.

\subsubsection{Silent Leave}
\label{sec:Silent leave}
Taking advantage of the relatively stable infrastructure in cloud environments, silent leave events are generally rare. We used the same heartbeat detection logic as SWIM \cite{das2002swim}. The superior performance of SWIM ensures that the failure detection mechanism maintains low performance overhead and detection latency, even as the cluster scales. In SWIM, each node periodically disseminates heartbeat messages to other nodes to verify their liveness. Importantly, both the expected time to first failure detection and the per-node message overhead remain constant, regardless of the cluster size. Silent node departures are promptly reported by other nodes, enabling the cluster to quickly converge to a correct and consistent state. Once such nodes are detected, the information is immediately broadcast to all other nodes using Reliable Message, ensuring that the cluster state is promptly restored to correctness. The default setting is to perform fault detection every second. In our tests, Snow consistently identifies faulty nodes within seconds and successfully completes their eviction. 

Even in the presence of broadcast failures, other nodes in the cluster will eventually detect the faulty node and report the fault. Unfortunately, during node failures, standard messages may be lost. Snow's broadcast path selection is dynamic, but in the standard Snow broadcast, each node has only one path to receive messages, resulting in potential message loss during brief failure windows. This problem has been proven to be unsolvable \cite{belsnes1976single} without delivering duplicates of a message. However, the Reliable Message feature ensures that messages are redelivered, guaranteeing the delivery of critical information. Newly joined nodes may initially miss these messages, yet through anti-entropy mechanisms, the cluster membership view can achieve eventual consistency. The system can recover within one minute \cite{dadgar2018lifeguard} even when multiple nodes fail simultaneously. This technique has been adopted in Consul \cite{hashicorp_consul}, which is widely used for member failure detection in production data center environments.
\subsection{Node Coloring}
\label{sec:Node Coloring}
\begin{figure}
    \centering
    \includegraphics[width=1\linewidth]{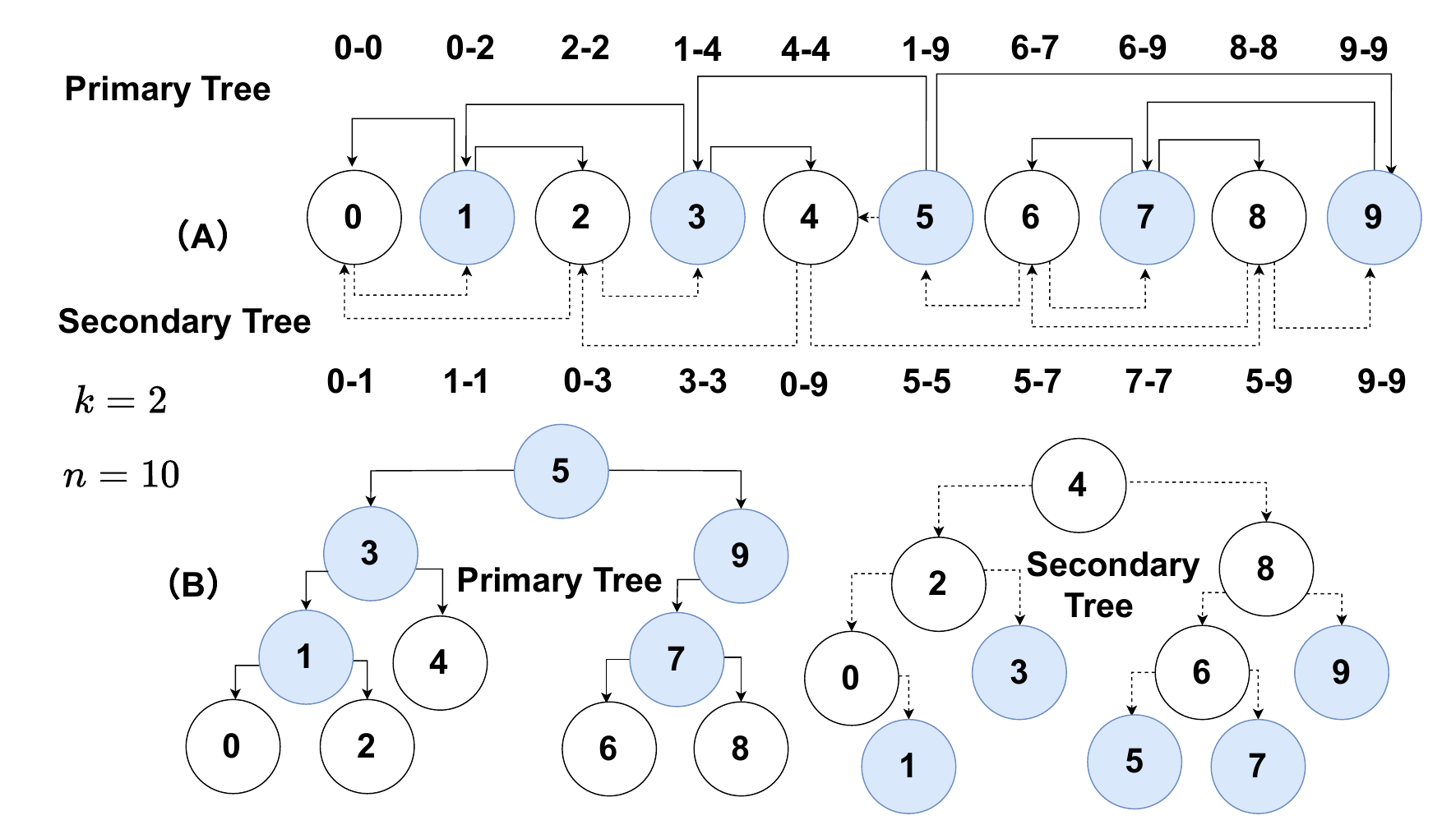}
    \caption{The solid line is the Primary Tree, and the dashed line is the Secondary Tree.}
    \label{fig:snow-coloring}
\vspace{-10pt}
\end{figure}
To improve the efficiency, stability, and propagation speed of message broadcasting in clusters, it is crucial to optimize the structure of the broadcast tree. Since nodes' join and leave are quickly broadcast to all nodes, most of the time, all nodes in the cluster have a consistent and correct view of the cluster. In this context, we observe the following issues: 1) Each node has only one parent node, which means that if the parent node fails or becomes a straggler, it significantly impacts the stability and performance of the cluster. 2) During a single message broadcast, the fan-out traffic of a node is greater than the fan-in traffic of the node. 3) Additionally, leaf nodes only receive data and do not send data. In the standard Snow algorithm, the broadcast tree is constructed on a ring, giving each node an equal probability of becoming a leaf node. However, when only a minority of nodes initiate transmissions, considerable bandwidth at the leaf nodes is underutilized.

Snow has implemented specific optimizations for scenarios where the cluster operates in a stable environment. In particular, we use some of the leaf nodes as internal nodes to construct a new tree. We refer to the original tree as the Primary Tree and the newly constructed tree as the Secondary Tree. The Secondary Tree broadcasts messages independently without consuming the fan-out bandwidth of internal nodes in the Primary Tree. This occurs because the leaf nodes of the Secondary Tree are always internal nodes of the Primary Tree.

Since this algorithm is an optimization triggered in a stable node scenario, we assume that all nodes have a correct, consistent, and immutable membership view during the execution process of broadcast. So, how does Snow construct the Secondary Tree? We rebuild a logical list based on the ring, which places the root node in the middle. As shown in Figure \ref{fig:snow-coloring} A. After that, partition the nodes into even and odd groups based on the parity of their indices in the array. During message dissemination, if the current node's index is odd, messages are preferentially sent to other odd-indexed nodes; If the index is even, messages are prioritized for even-indexed nodes. A node can send messages to a node with a different parity only if there are no nodes with the same parity within its assigned region (calculated separately for the left and right regions). We refer to this process as \textit{Node Coloring}. Figure \ref{fig:snow-coloring} shows how the trees are constructed. The above logical list is only used to distinguish the colors of nodes.

\textbf{Primary Tree:} As before, we logically place the broadcast origin at the middle of the ring and determine whether the root node is an even or odd node based on the number of nodes to its left. For instance, suppose the initial root node $N_0$ is an odd node, then $N_{-1}$ and $N_1$ must be even nodes. When identifying child node candidates in its region, the node prioritizes sending messages to other odd nodes within that region. If fewer than $k'$ odd nodes are available in that region, select even nodes instead for message transmission. This tree construction process can be viewed as two steps: First, a tree is constructed using odd nodes. Second, Using even numbered nodes as leaf nodes of the tree. Notably, even nodes are always restricted to leaf positions in the tree. The proof can be found in Appendix \ref{appendix:Leaf Node Proof}. Since even nodes are always leaf nodes, they increase the height of the Primary Tree by at most one level. This has a negligible effect on the cluster's convergence time. 

\textbf{Secondary Tree:} To construct the Secondary Tree, the initial node needs to send an additional message to node $N_{-1}$, which will serve as the root of the Secondary Tree. Therefore, the parity of the two root nodes must be different; furthermore, the initial boundaries of the root nodes of the two trees are the same. Thus, the height of the constructed Secondary Tree is similar to that of the Primary Tree. In practice, the two trees are constructed almost simultaneously and are both utilized for broadcasting. After the root node completes broadcasting to $k+1$ nodes, the Secondary Tree identifies its own root node, which is shown as node 4 in Figure \ref{fig:snow-coloring}.

According to the above design, we can prove that each node has two disjoint paths through which it can receive the message. For details, see Appendix \ref{appendix:Disjoint Paths Proof}. This evidence mitigates the straggler problem, as the proportion of straggler nodes is relatively small compared to the total number. Additionally, for each node, among the two paths, as long as one path successfully propagates the message, the node does not lose the message. As a trade-off, every node receives the same message twice, therefore, each node has at most two fan-in channels. In nodes with symmetric upstream and downstream bandwidth, the fan-in is 2 while the fan-out remains $k(k\geq 2)$. Obviously, when fan-in is less than or equal to fan-out traffic, the fan-in traffic does not become the bottleneck of the broadcast algorithm.

The Coloring messages still preserve the churn-tolerant property as proven in Appendix \ref{appendix:Convergence Proof}. This is because, for both the Primary Tree and the Secondary Tree, the only difference between their construction and that of the standard Snow algorithm's tree is that the selected nodes are either potential intermediate nodes or neighbor nodes of intermediate nodes. In addition to addressing the straggler problem, the Secondary Tree could also be used to optimize bandwidth utilization. Similarly to SplitStream \cite{castro2003splitstream}, data can be split and broadcast through both trees, fully leveraging the fan-out of each node. Due to space constraints, we do not elaborate further here. Nevertheless, SplitStream does not guarantee that messages sent to stable partial nodes are always reliable. The Node Coloring algorithm is not intended to replace our standard algorithm, but rather to serve as an additional optional feature. Using extra fan-out traffic from leaf nodes it accelerates cluster convergence and improves robustness. For scenes with large $k$, most nodes are leaves, making the Node Coloring algorithm particularly useful in such cases.

\begin{figure*}[t]
    \centering
    \includegraphics[width=1\linewidth]{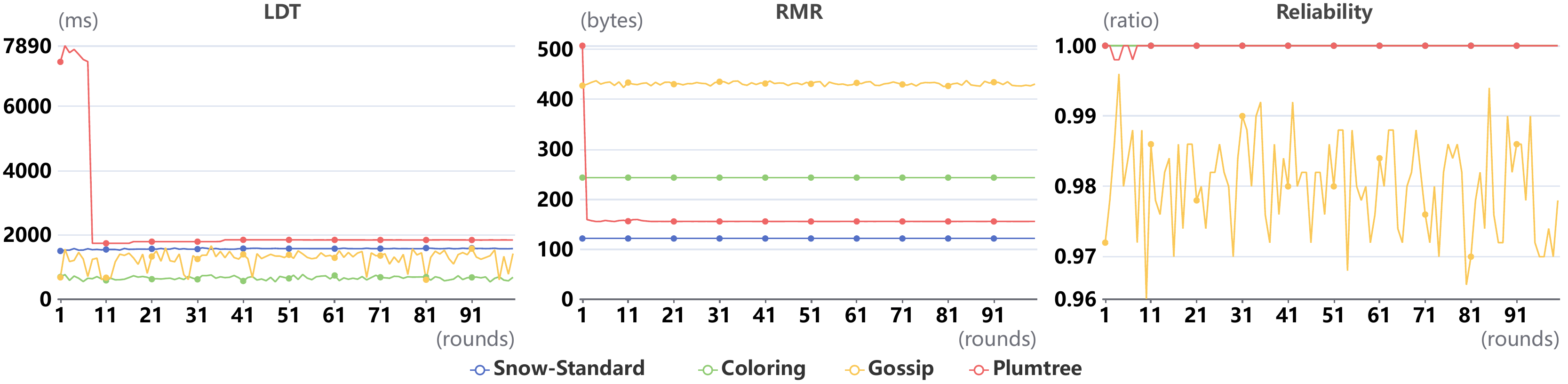}
    \caption{The x-axis denotes the sequence number of messages, where larger values correspond to messages sent at a later time.}
    \label{fig:stable}
    \vspace{-10pt}
\end{figure*}

\section{Experiment}
\label{sec:Evaluation}

\subsection{Metrics}
\label{sec:Metrics}
We define some metrics to evaluate the performance of broadcast protocols.

\textbf{Reliability: }This metric denotes the percentage of messages successfully delivered to the intended nodes. If Reliability reaches 100\%, the message dissemination is considered an atomic broadcast. It is worth noting that significant differences in Reliability render comparisons of other metrics invalid. Clearly, reliability is the most critical metric for message evaluation.

\textbf{Relative Message Redundancy(RMR): }RMR is used to measure the load of broadcast protocols. Defined as $\frac{m}{n}$. where $m$ is the total number of bytes received by all nodes from a single message, while a higher RMR value suggests poorer network traffic utilization. Lower Reliability results in a smaller RMR value. Therefore, the RMR protocol is only comparable under similar Reliability conditions. RMR is an important metric to measure the size of reactive message traffic and does not increase as the cluster scales up. This metric does not account for the messaging overhead incurred by the changes in the membership. This overhead is typically negligible and does not scale with the volume of messages.

\textbf{Last Delivery Time (LDT): } LDT refers to the time when the last node receives the message after the initial broadcast. We do not adopt Last Delivery Hop (LDH) because LDT better reflects the actual user experience. LDH represents the final hop in the message delivery path, whereas LDT captures the overall latency from the source to the destination, making it a more accurate measure of user-perceived performance. This metric captures the message convergence time, therefore, smaller values mean better performance.
\subsection{Experimental Setting}
\label{sec:Experimental Setting}
Within a cluster, each node forwards a received message to other nodes after a random assigned delay of 10–200 milliseconds, mimicking the inherent message processing time. To model straggler behavior, we designated 5\% of nodes as stragglers, which delay message forwarding by 1 second. All nodes were configured with a fanout $k$ of 4. At the beginning of the experiment, they had an accurate understanding of the state of the cluster.

Messages are broadcast at a fixed rate of one message per second. Each message is labeled by transmission order. The metrics derived from these messages provide insight into the evolution of the cluster's broadcast algorithm over time. We implemented these broadcast algorithms using a BIO communication model over TCP and set the fan-out of nodes to the same $k$ value. The experiments were conducted on a high-performance computing node equipped with a 24-core, 48-thread Intel Xeon 8269CY processor and 64GB of DDR4 RAM running at 2933 MHz. The software stack comprised Go 1.23 for algorithm implementation and an operating system with Ubuntu 22.04. We collected latency and packet loss data from 20 Alibaba Cloud servers within the same data center and utilized these metrics as network delay parameters. Leveraging Linux’s TC \cite{Almesberger_1999} command, we configured network parameters, allocating a 1MB buffer per port to emulate data center switches, with each node’s bandwidth capped at 1GB/s.

Currently, the vast majority of broadcast algorithms \cite{antunes_pulsarcast_2021,chu2002case} are designed for wide area networks, where the communication overhead between nodes is a critical parameter for optimization. However, this difference becomes negligible within data centers. We compare the implementations of the Gossip protocol, the standard Snow algorithm, the Coloring algorithm, and the Plumtree protocol \cite{leitao2007epidemic}. Among these, Gossip employs the simplest strategy: upon receiving a message, each node randomly forwards it to $k$ other nodes. This strategy is the most prevalent broadcast algorithm in data centers and has been widely adopted in numerous production systems \cite{akka,decandia2007dynamo}. Plumtree is also a decentralized broadcast protocol that does not account for communication latency between nodes and is used in some industrial scenarios \cite{vernemq} for local area network communication.

\subsection{Stable Environment}
In this experiment, no nodes will join or leave. Each node has an identical and accurate membership list. Figure \ref{fig:stable} presents LDT, RMR, and Reliability over 100 rounds of message transmission. Among them, the indicators of each message are sequentially numbered and displayed as the x-axis in the figure. The average values of all indicators are shown in Table \ref{tab:comparisonAtScenarios}.

Figure \ref{fig:stable} reveals that at the beginning of the broadcast, Plumtree exhibits severe fluctuations in terms of LDT, RMR, and Reliability as it relies on Gossip for initialization, which inevitably requires multiple rounds of message exchanges to establish an optimal path. In contrast, Snow avoids this issue by ensuring that the message forwarding path for each node remains deterministic. Additionally, due to the inherent design of Plumtree, the fan-out among nodes is unbalanced. This makes it difficult for users to configure a uniform fan-out value for Plumtree based on node bandwidth.

In Plumtree, the cluster requires a warming-up phase to construct the tree, when it starts or experiences considerable node changes before the LDT and RMR stabilize at an optimal level. The Plumtree is built with Gossip, and over time, all nodes eventually receive Gossip messages. However, message loss is still possible during initialization. Additionally, to determine the best topology, each node must maintain a set of child nodes for every root node. Consequently, whenever a new node initiates a broadcast as the root node, the cluster exhibits oscillations in RMR, LDT, and Reliability metrics, consistent with the patterns illustrated in the Figure \ref{fig:stable}. Although multiple nodes sharing the same tree is feasible \cite{kaneko2023broadcast}, these trees are suboptimal for all nodes except the root (the messages shown in Figure \ref{fig:stable} originate from the same node). Therefore, Plumtree performs well only when a small number of nodes in the cluster need to broadcast messages in a stable environment.
% In contrast, Snow does not require a warming-up phase, each forwarding node simply follows predefined rules for message forwarding without considering the message source (i.e., the root node). 
% In comparison, when a message is first broadcast, Snow’s LDT, RMR, and reliability do not exhibit oscillations.

Although each node in a Gossip protocol has the same fan-out value, a substantial amount of traffic is wasted by broadcasting messages to nodes that have already received them, as illustrated by the RMR in Figure \ref{fig:stable}. Furthermore, as discussed in Section \ref{sec:Existing Protocols}, if the $k$ does not scale with the cluster size, Gossip is highly likely to fail in delivering messages to all nodes. The LDT of the Gossip protocol is not particularly prominent. However, we emphasize that it never achieves 100\% Reliability, which in turn makes other performance metrics appear more favorable. 

In Snow, the RMR of the Node Coloring algorithm is twice that of the standard algorithm. Nevertheless, during a single independent broadcast, each node forwards the message to at most $k$ nodes (except for the root node, which forwards to $k+1$ nodes). As mentioned in Section \ref{sec:Node Coloring}, this additional traffic originates entirely from leaf nodes, without consuming the fan-out capacity of internal nodes. The Node Coloring algorithm constructs two trees, which brings additional benefits. As previously mentioned, each node has two distinct paths to receive messages. This partially mitigates the straggler problem and speeds up message convergence. Furthermore, since both trees are built simultaneously, when message convergence occurs, both trees might have yet to complete their respective propagations. This characteristic contributes to optimizing the delivery time.

\begin{table*}[t]
\centering
\begin{tabular}{l ccc ccc ccc}
\toprule
\textbf{AVG} & \multicolumn{3}{c}{\textbf{LDT (ms)$\downarrow$}} & \multicolumn{3}{c}{\textbf{RMR (bytes)$\downarrow$}} & \multicolumn{3}{c}{\textbf{Reliability (ratio)$\uparrow$}} \\
       $n=500,k=4$     & Stable & Churn & Breakdown & Stable & Churn & Breakdown  & Stable & Churn & Breakdown\\
\cmidrule(lr){2-4} \cmidrule(lr){5-7} \cmidrule(lr){8-10}
Gossip                     & 1608       & 1278        & 1250         & 432           & 432        & 428       & 0.954     & 0.950      & 0.971 \\
Plumtree                   & 3183       & 8099        & 4588         & 160           & 184        & 160       & 0.999     & 0.998      & 0.990 \\
\textbf{Snow-Standard}     & 1560       & 1561        & 1598         & \textbf{122}  &\textbf{122}&\textbf{121}& \textbf{1}& \textbf{1}& 0.990     \\
\textbf{Coloring}          &\textbf{652}&\textbf{634} & \textbf{760} & 244           & 244        & 241       & \textbf{1}& \textbf{1} &\textbf{0.991}\\

\bottomrule
\end{tabular}
\caption{Performance Comparison Across Different Scenarios.}
\label{tab:comparisonAtScenarios}
    \vspace{-8pt}
\end{table*}

\begin{figure}[t]
    \centering
    \includegraphics[width=1\linewidth]{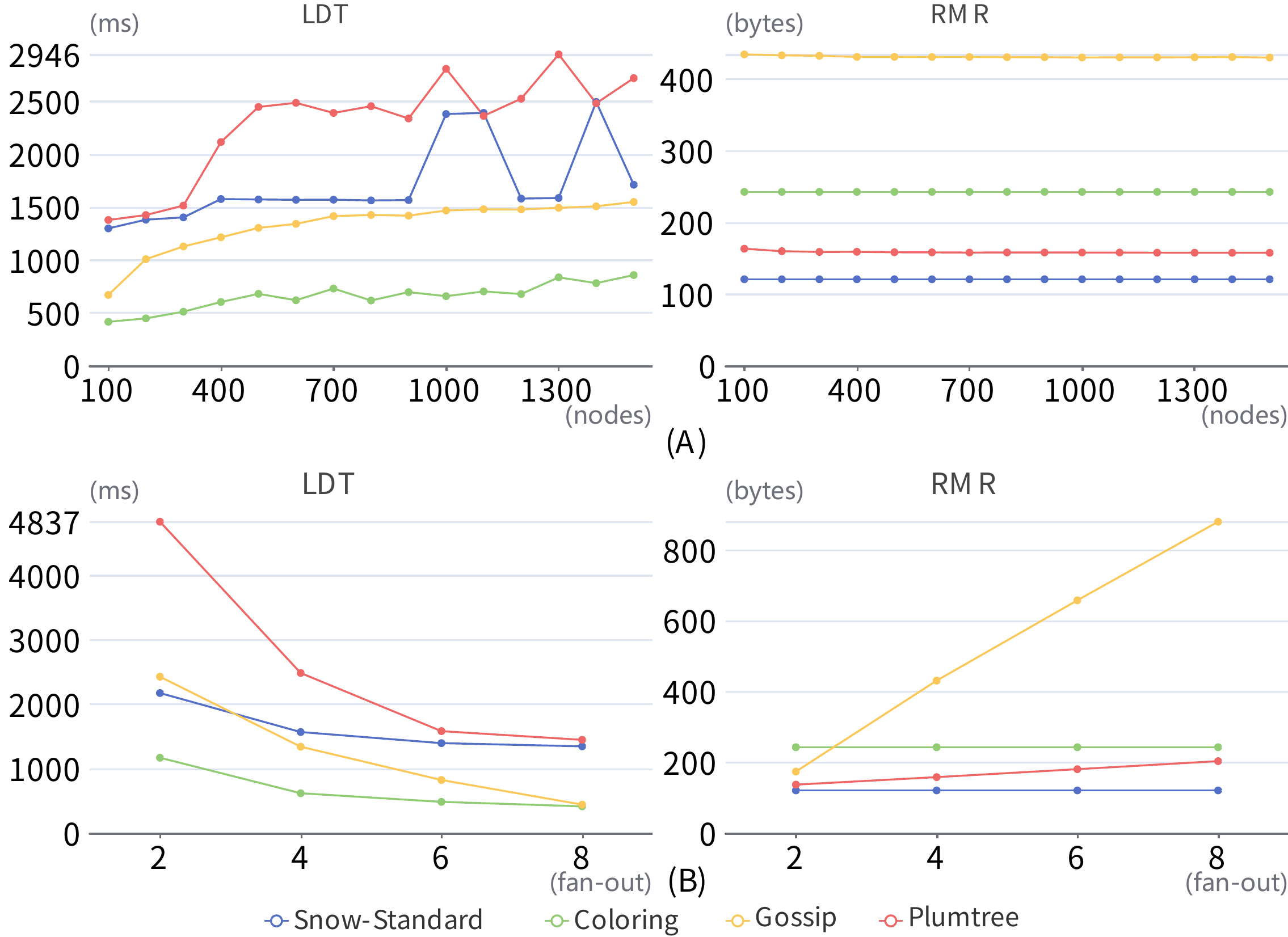}
    \caption{A shows the adjustment of $n$ with fixed $k=4$; B shows the fixed $n=600$ adjusted $k$.}
    \label{fig:change num and k}
    \vspace{-10pt}
\end{figure}

\begin{figure*}[t]
    \centering
    \includegraphics[width=1\linewidth]{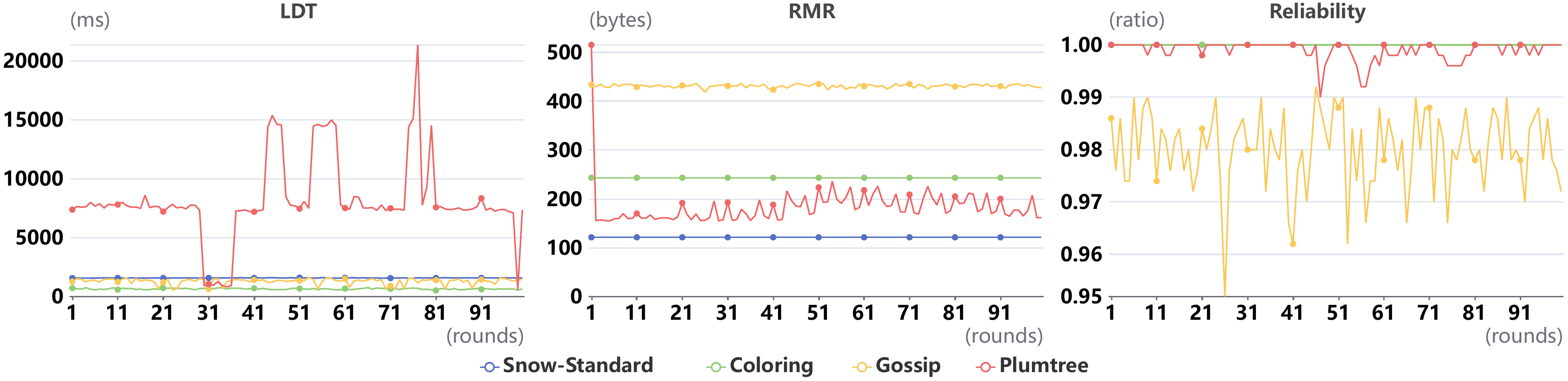}
    \caption{With 500 fixed nodes, the figure shows the metrics for a fixed subset of nodes under frequent node churn in the cluster.}
    \label{fig:leaving and joining}
    \vspace{-10pt}
\end{figure*}

\textbf{When $k$ and $n$ take various values:} In real-world environments, the number of nodes is unfixed. To investigate the impact of node count variations on LDT and RMR, we fix the $k$ at 4 and gradually increase the number of nodes $n$ from 100 to 1500. The results are shown in Figure \ref{fig:change num and k}A, which illustrates the effect of a linear increase in the number of nodes on cluster LDT and RMR. In tree-based broadcast algorithms, an increase in cluster size does not directly lead to an increase in LDT. Instead, LDT only increases when the growth in cluster size affects the tree height. As seen at $n=500$ to $n=900$, since the increase does not alter the tree height, the cluster's LDT remains unchanged. Additionally, for Snow and Gossip, the RMR value does not increase with the total number of nodes. In the figure, we observe that due to Plumtree’s initial bias in broadcasting, messages could persistently propagate along suboptimal paths. If this occurs while the cluster remains stable, the suboptimal path could continue to be used indefinitely. Overall, the delivery time of the standard Snow algorithm is on average over 30\% faster than Plumtree. Oscillations in the standard Snow algorithm emerge when straggler nodes become internal nodes, thereby increasing message propagation latency to descendant nodes. Our Coloring algorithm addresses this limitation by establishing disjoint message propagation paths.

The fan-out value $k$ is a user-adjustable parameter. In our experiments, we fixed the number of nodes at 600 and gradually increased $k$ from 2 to 8. Figure \ref{fig:change num and k}B illustrates the impact of the $k$ on LDT, indicating that the optimization effect of increasing the $k$ tends to saturate when the value becomes too large. This phenomenon arises as a reduction in LDT only occurs when increasing the fan-out, which effectively reduces the tree height. 
% Conversely, if a higher $k$ no longer decreases the tree height, each node will have to forward more messages, which instead leads to an increase in LDT.
Moreover, we observed that in the Snow protocol, the RMR of nodes remains stable regardless of changes in the $k$. This occurs because RMR measures the average fan-out traffic per node. As the fan-out increases, more nodes become leaf nodes, which have a $k$ of zero, thereby keeping the overall average fan-out stable.

\begin{figure*}[t]
    \centering
    \includegraphics[width=1\linewidth]{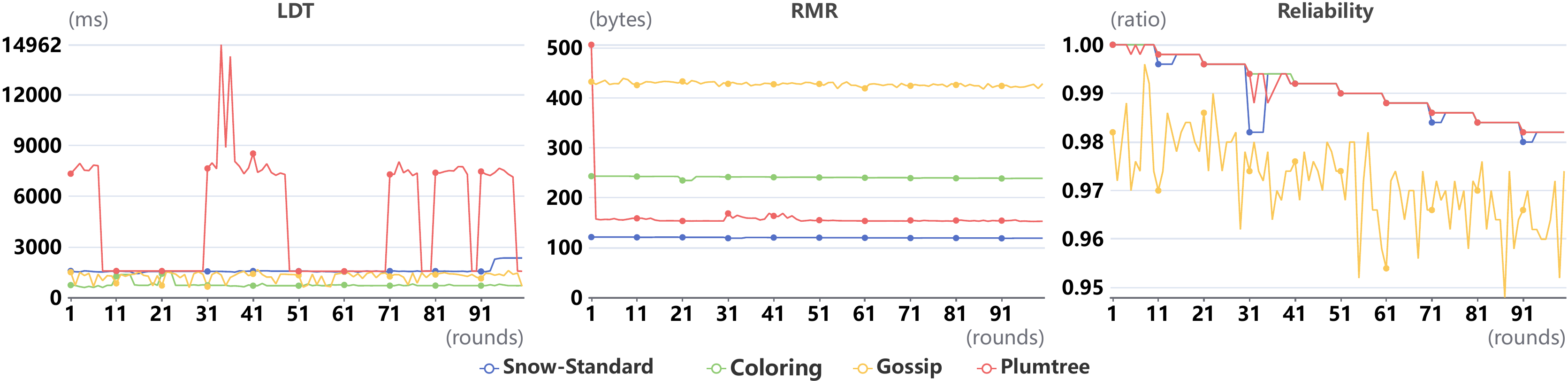}
    \caption{With 500 fixed nodes, for every 10 messages sent, one node silently leaves.}
    \label{fig:eval-Breakdown}
    \vspace{-10pt}
\end{figure*}

In Snow, the tree height determines the number of hops required for message delivery, where each hop corresponds to a message forwarding event. The relationship between tree height, $n$, and $k$ is as follows:
\begin{equation}
H = \left\lceil \log_k \left( (k-1)n \right) + 1 \right\rceil
\end{equation}
Here, H represents the tree height. We observe that as $n$ increases, the tree height grows logarithmically, which is consistent with our experimental data. Snow users are enabled to calculate the time required for message convergence based on their chosen values of $k$ and $n$, as well as the time required for message convergence.

\subsection{Frequent Leaving and Joining}
In a cloud environment, the dynamic joining and leaving of nodes are common occurrences. Yet we do not want changes in certain parts of the cluster to affect the already stable portions. Therefore, we designed the following experiment. Similarly, we initialize 500 nodes with correct membership view. Subsequently, we launch a new node and integrate it into the cluster. The node’s joining process occurs concurrently with message broadcasting. After transmitting 10 messages, we gracefully decommission the new node. This scenario is common in data centers, where node departures and arrivals occur while other nodes remain operational. Crucially, neither node joining nor leaving should disrupt the stable portion of the cluster. To evaluate the impact of the newly joined and departed nodes, we collect metric data exclusively from the fixed (500) nodes, ensuring that our analysis only focuses on stable nodes.

Figure \ref{fig:leaving and joining} illustrates the significant difference in Plumtree's Reliability between stable conditions and scenarios with frequent node churn. This discrepancy is attributed to Plumtree's difficulty in maintaining a stable topology under high churn. Each node's arrival or departure necessitates partial reconstruction of the tree, leading to a substantial decline in Plumtree's overall stability. For comparison, the two algorithms proposed in Snow leverage a robust fault tolerance model, guaranteeing that the fixed subset of nodes reliably retains all messages. Consequently, our protocol exhibits consistently low fluctuations in LDT, demonstrating its resilience under churn conditions.

% In contrast, Snow maintained a balanced multi-way tree as the final dissemination topology, but individual nodes do not need to explicitly construct such a tree. As a result, node churn does not affect the stable portion of the cluster. Since all nodes in the cluster maintain an accurate view of the stable nodes (with new nodes synchronizing existing states upon joining), our broadcast model ensures that the stable nodes do not experience message loss.

\subsection{Breakdown}

Although hardware or system-level crashes are relatively infrequent compared to regular node joins and graceful departures, we nonetheless evaluate our system under such failure scenarios. In this case, nodes do not undergo a graceful shutdown but instead silently leave the system due to abrupt termination. To emulate node failures, we drop both inbound and outbound traffic of a randomly selected node after every 10 messages are sent. Before other nodes detect a node crash, the new 10 messages will continue to be sent. For external observer nodes, this behavioral pattern is entirely indistinguishable from a node failure scenario. External nodes cannot differentiate whether the node has actually departed or the network has become unreachable \cite{chandra1996unreliable}. Compared to a kill operation, this approach more accurately simulates extreme crash scenarios where the operating system is unable to actively terminate TCP connections with other nodes.

Figure \ref{fig:eval-Breakdown} shows the scene of nodes silently leaving. Due to partial node failures, after a machine malfunction, Reliability never reaches 100\%. In the case of Snow’s standard algorithm, the final broadcast path will form a tree, leads to potential disconnections under bursty failures, causing downstream nodes to miss messages entirely. Plumtree employs Gossip-based recovery during churn, resulting in Reliability fluctuations with each membership change. In contrast, the Coloring algorithm maintains two disjoint paths for message delivery at every node, providing resilience to failure and achieving the highest Reliability across all scenarios, as reflected in the Breakdown metric in Table \ref{tab:comparisonAtScenarios}. Theoretically, under frequent cluster membership changes, the Coloring algorithm may still permit message loss among healthy nodes due to inconsistent membership views across the cluster, albeit with low probability. Nevertheless, across all our experimental trials, we have never observed this phenomenon occurring. As with previous observations, Plumtree exhibits significant instability in LTD under high churn, whereas both the Coloring algorithm and Snow’s standard protocol are unaffected by frequent membership changes. Figures \ref{fig:leaving and joining} and \ref{fig:eval-Breakdown} demonstrate characteristic performance patterns of Plumtree in unstable environments. Across all test scenarios, our Coloring algorithm consistently achieved optimal performance metrics except for RMR. The observed RMR overhead stems exclusively from leaf nodes, making this optimization effectively cost-free when broadcasts are initiated by a single node or a small subset of the cluster.

\section{Conclusion}
In this paper, we propose a new algorithm for message broadcast. We designed Snow based on a key insight: In a cloud or data center environment, the duration of cluster stability significantly exceeds the frequency of disruptive changes. Furthermore, Normal membership changes occur far more frequently than failures that cause unexpected node departures. To exploit these characteristics, Snow employs targeted optimizations. During membership changes, it preserves the stability of unaffected nodes. When membership changes occur, Snow ensures that stable nodes remain unaffected, even if the global view becomes temporarily inconsistent. In cases of unexpected node failures, we guarantee the delivery of critical messages.

Compared to existing algorithms, we not only propose a churn-tolerant model but also demonstrate in our experiments that our model achieves higher reliability and efficiency. Our designed algorithm differs from all previous approaches. Under the standard Snow algorithm, routine node joins and departures do not affect the QoS of message broadcasting, and each node receives the message only once. With the Node Coloring algorithm, Snow efficiently utilizes the bandwidth of leaf nodes, addressing both broadcast failures caused by abnormal node failures and performance degradation due to straggler nodes.

\newpage
\bibliographystyle{plain}
\bibliography{bibs}
\appendix

\section{Convergence Proof}
\label{appendix:Convergence Proof}
\textbf{Theorem}: If all nodes in the cluster know a node and the message is not lost, the probability of this node receiving the message is 100\%.

\textbf{Proof}: We define $N_x$ as a node knows to all nodes, where $N_x$ refers to any node in the set $S=\{N_t | t \in [1:n]\}$, excluding the root node $N_0$. The root node $N_0$ possesses the set $S$. Given that all nodes are aware of node $N_x$, it follows that $N_x$ is necessarily contained in the set $S$. Each node partitions $S$ into $k$ subsets of approximately equal size. According to our design, $N_x$ must reside in one of the regions. We assume it belongs to the $i$-th region, defined as:
\begin{equation}
S_i= \{N_t | t \in [\frac{(i-1) n'}{k'}+1 , \frac{i n'}{k'}]\},\quad N_x \in S_i
\end{equation}
The intermediate node of the $i$-th part is $N_{(\frac{(2i-1) n'}{k'}+1)/2}$, if $(\frac{(2i-1) n'}{k'}+1)/2=x$, $N_x$ receives the message, as the current node disseminates the message to the corresponding intermediate nodes of the $k$ regions during the broadcasting process. Otherwise, treat $S_i$ as the next set to be processed, and repeat the above operations until $N_x\in S_{i...j}\wedge|S_{i...j}|=1$. At this stage, $N_x$ is the only node in the region; Therefore, it must receive the broadcast from the parent node. In conclusion, $N_x$ is guaranteed to receive the message.

\section{Broadcast With Different Membership}
\label{appendix:diff}
\begin{figure}[htbp]
    \centering
    \includegraphics[width=1\linewidth]{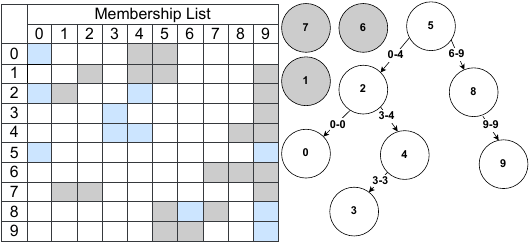}
    \caption{Broadcast situation when $n=10$ and Membership are not synchronized. 
Each row along the way is a membership view of a node.}
    \label{fig:diff-member}
\end{figure}
Figure \ref{fig:diff-member} illustrates the broadcasting process when each node maintains a distinct membership view. Each row in the figure represents the state held by a particular node, where grey cells indicate nodes that are absent from the current membership view. Blue cells denote boundary nodes for each node. If no boundary nodes exist, they are inserted into the membership view before the computation begins—thus, boundary nodes are always included in the membership list. When there is only one boundary node, it must be a leaf node. In the figure, nodes 3 and 0 are known to all nodes and are therefore guaranteed not to lose any information (The columns of 0 and 3 are either blue or white). When a node is only partially known within the cluster, it may or may not receive broadcast messages depending on its connectivity. For instance, nodes 8 and 9 can successfully receive broadcasts, whereas nodes 1, 6, and 7 fail to acquire the transmitted messages.

\section{Leaf Node Proof}
\label{appendix:Leaf Node Proof}
\textbf{Theorem}: If the root node is an odd node, then every even node is a leaf node.

\textbf{Proof}: We prove this theorem by contradiction. It is known that odd nodes are always selected before even nodes. Assume, for the sake of contradiction, that there exists an even node $v$ whose parent node $u$ is an odd node. By the design of the Snow selection strategy, given that the nodes added to the tree first are odd, if an even node that is not a leaf exists, then such a node $v$ must exist. Since $v$ is not a leaf, it has at least one child node. Therefore, the region assigned to $u$ by its parent node must contain at least two even nodes(note that $v$ counts as one even node in the region). The only way for the region assigned to $u$ to contain at least two nodes is to contain strictly more than $k'$ even nodes. However, according to Snow's search strategy, odd nodes are always selected before even nodes, and the region must contain at least $k'-1$ nodes. Consequently, only one even node would eventually be processed. This contradicts the earlier conclusion that the existence of two nodes within the region of node $v$. In the same manner, we obtain that if the root node is an even node, then every odd node is a leaf node.

\section{Disjoint Paths Proof}
\label{appendix:Disjoint Paths Proof}
\textbf{Theorem}: For any node in the stable cluster, the Coloring algorithm generates two disjoint access paths.

\textbf{Proof}: Assume the set of all nodes in the cluster is denoted by $S$. We define the internal nodes and leaf nodes of the primary tree as $S_i^p$ and $S_l^p$, respectively. Similarly, the internal and leaf nodes of the secondary tree are denoted by $S_i^s$ and $S_l^s$. Furthermore, we have:

\begin{equation}
S_i^p+S_l^p=S_i^s+S_l^s=S,\quad
S_i^p \cap S_l^p = \emptyset , S_i^s \cap S_l^s = \emptyset
\end{equation}
For any node $N_x$, the set of nodes traversed from the root of the primary tree to $N_x$ is denoted as $S_r^p$, and the corresponding set in the secondary tree is denoted as $S_r^s$. In other words, we aim to prove that:
\begin{equation}
S_r^p \cap S_r^s= \emptyset
\end{equation}
Since only internal nodes are traversed along the path, we have:
\begin{equation}
S_r^p\subseteq S_i^p,S_r^s\subseteq S_i^s
\end{equation}
According to the proof in Appendix \ref{appendix:Leaf Node Proof}, the internal nodes of one tree must be the leaf nodes of another tree. We know that:
\begin{equation}
S_i^p \subseteq S_l^s,S_i^s \subseteq S_l^p
\end{equation}
Therefore, we conclude that: $S_i^p \cap S_i^s = \emptyset$, which implies: $S_r^p \cap S_r^s= \emptyset$. 

\clearpage
\appendix

\end{document}